\documentclass[prl,aps,reprint,amsmath,amssymb,floatfix,superscriptaddress]{revtex4-2}

\usepackage[utf8]{inputenc}
\usepackage[colorinlistoftodos]{todonotes}
\usepackage[colorlinks,citecolor=blue,linkcolor=blue]{hyperref}
\usepackage{graphicx}	
\usepackage{dcolumn}	
\usepackage{bm}			
\usepackage{siunitx}    
\usepackage{color,soul}

\newcommand{\CNN}{Centre de Nanosciences et de Nanotechnologies, CNRS, Universit{\'e} Paris-Saclay, 91120 Palaiseau, France}
\newcommand{\MPG}{Max Planck Institute for the Science of Light, Staudtstra{\ss}e 2, 91058 Erlangen, Germany}
\newcommand{\LMOPS}{Laboratoire Matériaux Optiques, Photonique et Systèmes, CentraleSup{\'e}lec, Universit\'{e} de Lorraine, 57070 Metz, France}

\begin{document}

\title{Analytical model of the inertial dynamics of a magnetic vortex}

\author{Myoung-Woo Yoo}
\affiliation{\CNN}
\affiliation{\LMOPS}
\author{Francesca Mineo}
\affiliation{\CNN}
\affiliation{\MPG}
\author{Joo-Von Kim}
\affiliation{\CNN}

\date{\today}

\begin{abstract}
We present an analytical model to account for the inertial dynamics of a magnetic vortex. The model is based on a deformation of the core profile based on the D{\"o}ring kinetic field, whereby the deformation amplitudes are promoted to dynamical variables in a collective-coordinate approach that provides a natural extension to the Thiele model. This extended model accurately describes complex transients due to inertial effects and the variation of the effective mass with velocity. The model also provides a quantitative description of the inertial dynamics leading up to vortex core reversal, which is analogous to the Walker transition in domain wall dynamics.
\end{abstract}

\maketitle

Topological solitons represent compact, nontrivial solutions of a nonlinear field system. In magnetic thin films, examples of such excitations include one-dimensional (1D) configurations like domain walls~\cite{Bloch:1932ux, Hubert1998} and two-dimensional structures such as vortices~\cite{Gouvea:1989gb, Papanicolaou:1991eg, Cowburn1999, Shinjo2000} and skyrmions~\cite{Bogdanov:1994bt, Bogdanov:1989vt}.
For the latter, a common approach for describing their dynamics involves a collective coordinates approach in which the position $\mathbf{X} = (X, Y)$ of the vortex or skyrmion core, assumed to be rigid, is elevated to a dynamical variable $\mathbf{X}(t)$ and subsequently defines the entire dynamics of the system by allowing all other degrees of freedom to be integrated out. This is the basis of the Thiele model~\cite{Thiele1973, Huber1982, Papanicolaou:1991eg},
\begin{equation}
		\mathbf{G} \times \dot{\mathbf{X}} + \alpha \tensor{D} \cdot \dot{\mathbf{X}} = -\frac{\partial U}{\partial \mathbf{X}},
\label{Eq:Thiele}
\end{equation}
where $\mathbf{G} = (M_\mathrm{s}/\gamma)\int dV \sin\theta \left( \nabla \theta \times \nabla \phi \right)$ is the gyrovector, $\tensor{D} = (M_\mathrm{s}/\gamma)\int dV \left[ \nabla \theta \otimes \nabla \theta + \sin^2\theta \left( \nabla \phi \otimes \nabla \phi \right)  \right]$ is the damping tensor, $\alpha$ is the Gilbert constant,  $U$ is a total magnetic energy, $M_\mathrm{s}$ is the saturation magnetization, and $\gamma$ is the gyromagnetic constant. Here, $\theta$ and $\phi$ represent the orientation of the magnetization field, $\mathbf{m}$, in spherical coordinates.

However, it has been known since the seminal work of D{\"o}ring~\cite{Doring1948} that soliton motion in ferromagnetic materials should be accompanied by a deformation of the core, since the Landau-Lifshitz equation governing the magnetization dynamics does not exhibit Galilean invariance and so changes to the static profile must appear for a soliton propagating at finite velocity in the steady state. This is apparent in the 1D model of domain wall dynamics, where in addition to the wall position $q(t)$ an additional dynamical variable, the \emph{wall angle} $\varphi(t)$, appears in the equations of motion and describes a transformation from the equilibrium static wall profile~\cite{Slonczewski1972, Schryer1974}. In addition, the potential energy related to $\varphi(t)$ characterizes the wall mass and limits the propagation speed of the time-independent profile, a phenomenon known as Walker breakdown.

Interestingly, despite the large body of theoretical work on magnetic vortex and skyrmion dynamics to date, a consistent description of inertial effects and core deformation within the Thiele framework remains largely unexplored. Some earlier work has identified the importance of deformation~\cite{Ivanov:1989fx, Ivanov:1998bm, Zagorodny:2004ch}, but appears to be overlooked. From a phenomenological standpoint, inertia can be introduced through the addition of a mass term $\mathcal{M}$~\cite{Volkel:1994en, Wysin1996, Mertens:1997jr, Ivanov:1998bm, Sheka:2001kz, Kovalev:2003eh, Sheka:2006iw, Guslienko2006}, 
\begin{equation}
\mathcal{M} \ddot{\mathbf{X}} +	\mathbf{G} \times \dot{\mathbf{X}} + \alpha \tensor{D} \cdot \dot{\mathbf{X}} = -\frac{\partial U}{\partial \mathbf{X}}.
\label{Eq:MassThiele}
\end{equation}
However, this approach neglects the change in internal energy as a result of changes to the magnetic configuration of the core, which occurs in processes such as vortex core switching~\cite{Hertel:2006hn, VanWaeyenberge2006, Yamada2007, Guslienko2008} and trochoidal motion of antiskyrmions~\cite{Ritzmann2018}, which are phenomena analogous to Walker breakdown.

In this Letter, we develop a model to account for the motion-induced inertia of magnetic vortices. First, we describe an analytic model for the deformation based on D{\"o}ring's kinetic field, then derive equations of motion of the vortex core, from which the inertial motion can be described. We define the effective inertial mass of the vortex core, and show that the mass depends on the core velocity. We propose a simple model for the energy profile which can be exploited to understand the core's inertial motion, dependence of the mass on the velocity, and dynamics of the core before the core polarity switching.

\begin{figure}[t]
\centering\includegraphics[width=8.6cm]{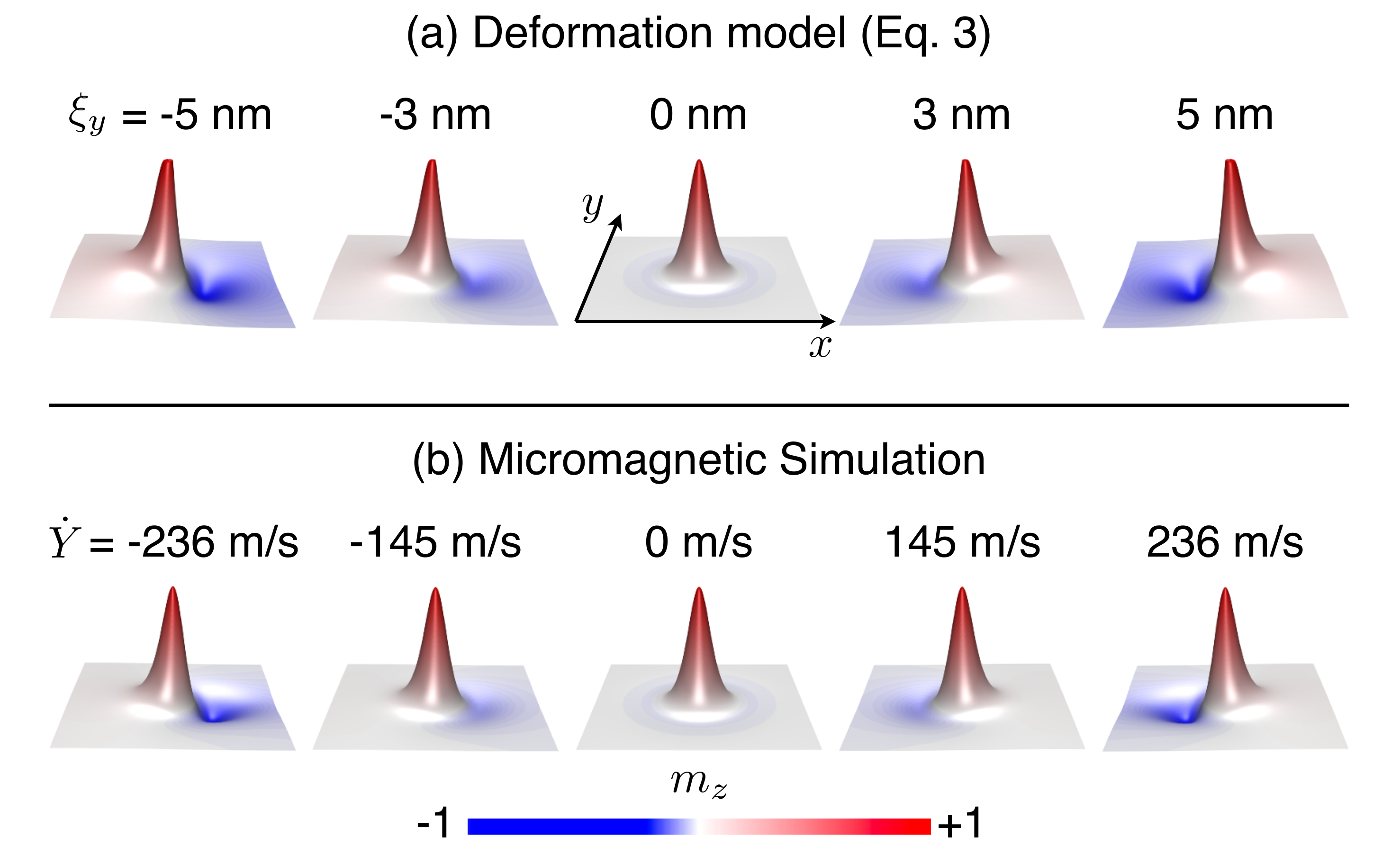} 
\caption{\label{Fig:Deformation}
 (a) Deformed vortex configurations near the core ($60 \times 60 \text{ nm}^2$) calculated from the static configuration ($\xi_x = 0 \text{ nm}$) using Eq.~(\ref{Eq:config}) with $\xi_x = -5, -3, 3, \text{ and } 5 \text{ nm}$. The height is proportional to $m_z$ and color indicates $m_z$ as noted by the color bar. (b) Deformation of the vortex obtained from a micromagnetic simulation at the indicated velocity, $\dot{Y}$.
   }
\end{figure}

Consider a moving vortex with a finite velocity, $\mathbf{\dot{X}}$. In this case, the magnetic configuration of the vortex should be deformed by D{\"o}ring's kinetic field, $\mathbf{H}_\mathrm{kin} = (1/\gamma_0) \mathbf{m} \times \left[ (\dot{\mathbf{X}} \cdot \bm{\nabla}) \mathbf{m}\right]$, where $\gamma_0 = \mu_0 \gamma$ and $\mathbf{m} = (\sin \theta \cos \phi, \sin \theta \sin \phi, \cos \theta)$ is a unit vector of a local magnetic moment \cite{Doring1948, Yamada2007}. $\theta$ and $\phi$ are polar and azimuthal angles of the magnetic moment. Our model is based on the assumption that deformations $\delta \mathbf{m}$ to the static core profile, $\mathbf{m}_0$, can be expressed in terms of a rotation of the magnetic moments towards the direction of the kinetic field, i.e., $\delta \mathbf{m} \propto -\mathbf{m}_0 \times \left( \mathbf{m}_0 \times \mathbf{H}_\mathrm{kin}\right)$~\cite{Troncoso:2014gt}. This leads to the deformation \emph{ansatz},
\begin{subequations}
\begin{align} 
		\theta(\mathbf{x}, \bm{\xi}) &= \theta_0 -\sin \theta_0 \left[ \bm{\xi} \cdot \bm{\nabla} \phi_0 \right]; \\
		\phi(\mathbf{x}, \bm{\xi}) &= \phi_0 +\frac{1}{\sin \theta_0} \left[ \bm{\xi} \cdot \bm{\nabla} \theta_0 \right],
\end{align}
\label{Eq:config}
\end{subequations}
where $\theta_0 = \theta_0(\mathbf{x}-\mathbf{X})$ and $\phi_0 = \phi_0(\mathbf{x}-\mathbf{X})$ are the polar and azimuthal angles of the static configuration. $\bm{\xi}(t) = (\xi_x(t), \xi_y(t))$ is a new dynamical variable that describes the deformation amplitude with dimensions of length. Using Eq.~(\ref{Eq:config}), we obtain the deformed configuration from the initial vortex state as presented in Fig.~\ref{Fig:Deformation}(a) in the case of a permalloy disk with a diameter of 512 nm and a thickness of 20 nm. The result shows that Eq.~(\ref{Eq:config}) gives largely good description of the core deformation observed in a micromagnetic simulation [Fig.~\ref{Fig:Deformation}(b)], in particular, its asymmetric dip formation near the vortex core~\cite{VanWaeyenberge2006, Vansteenkiste2009, Vansteenkiste2014}.

From Eq.~(\ref{Eq:config}), equations of motion can be derived from the Euler–Lagrange equations~\cite{Slonczewski1972, Thiaville2006, Kim2012, SM} of the
%
%
four independent collective coordinates, $X$, $Y$, $\xi_x$, and $\xi_y$, %
\begin{subequations}
\begin{align} 
	\mathbf{G} \times \left( \dot{\mathbf{X}} - \alpha \dot{\bm{\xi}} \right) +  \tensor{D} \cdot \left( \alpha \dot{\mathbf{X}} + \dot{\bm{\xi}} \right) &= -\frac{\partial U}{\partial \mathbf{X}}; \\
	\mathbf{G} \times \left( \dot{\bm{\xi}} + \alpha \dot{\mathbf{X}} \right) + \tensor{D} \cdot \left( \alpha \dot{\bm{\xi}} - \dot{\mathbf{X}} \right) &= -\frac{\partial U}{\partial \bm{\xi}},
\end{align}
\label{Eq:extThiele}
\end{subequations}
where the Usov ansatz of the vortex core profile leads to $\mathbf{G} =  2 \pi M_\mathrm{s} p d / |\gamma| \, \hat{\mathbf{z}}$ and $\tensor{D} = D \tensor{I} = \tensor{I} (\pi M_\mathrm{s} d / |\gamma|) \left[ 2 + \ln(R/b) \right]$, with $\tensor{I}$ being the $2\times 2$ identity matrix~\cite{Guslienko2006}. Here, $d$ is the film thickness, $p$ is the core polarity, $R$ is the system size, and $b$ is the vortex core radius.

\begin{figure}
\centering\includegraphics[width=6.5cm]{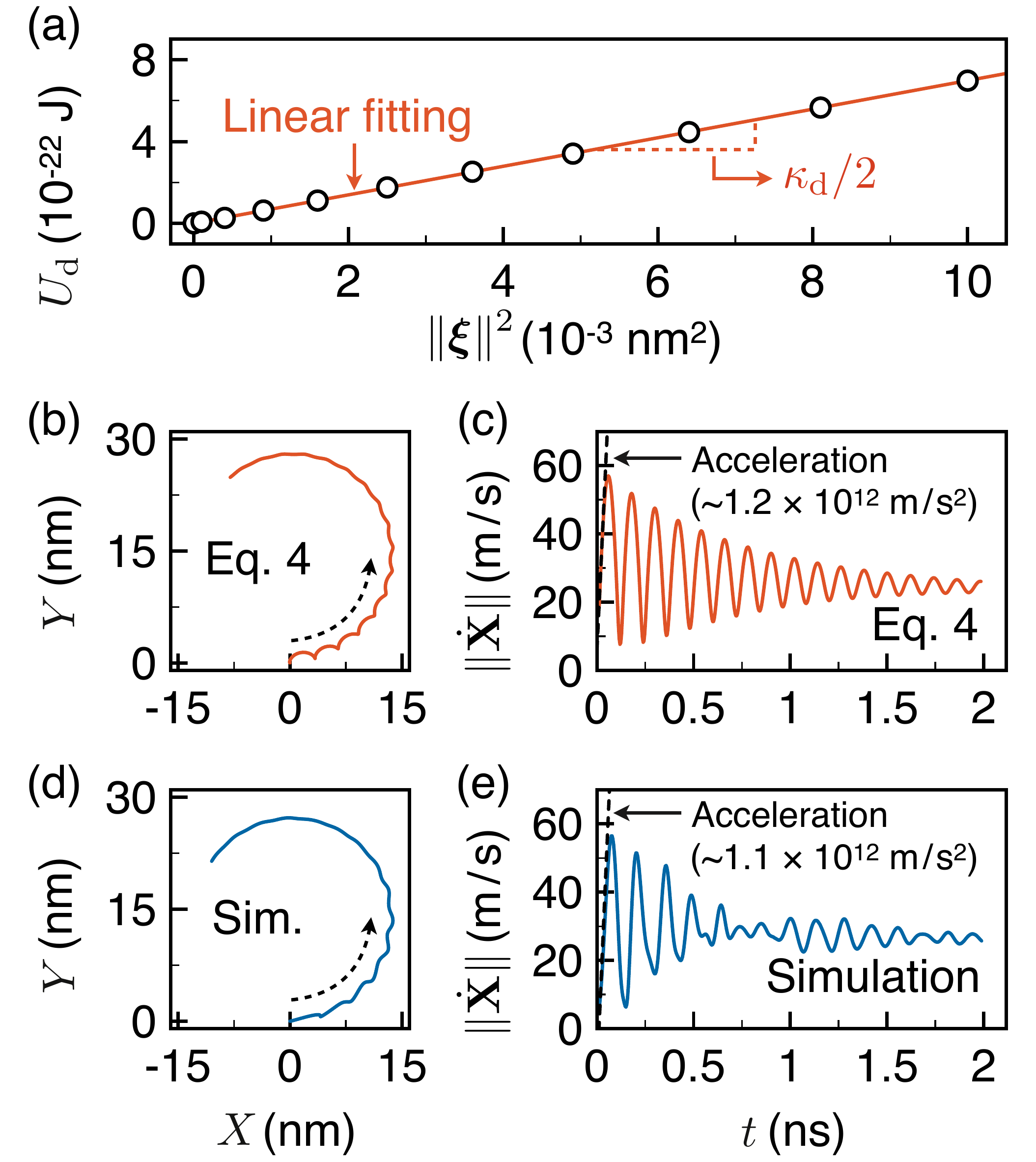} 
\caption{\label{Fig:kd}
	(a) Deformation energy, $U_\mathrm{d}$, as a function of $\| \bm{\xi} \|^2$ (symbols) with the linear fit (red line). (b) Trajectory of the vortex core for 2 ns obtained by solving Eq.~(\ref{Eq:extThiele}) with $\kappa_\mathrm{d} = 0.14 \text{ J} / \text{m}^2$. The motion is excited by an in-plane 2 mT magnetic field. (c) Time evolution of the core speed, $\| \dot{\mathbf{X}} \|$, during the motion in (b). (d) Core trajectory obtained from micromagnetic simulation. (e) Time evolution of $\| \dot{\mathbf{X}} \|$ during the motion presented in (d). The dashed lines in (c) and (e) indicate the initial acceleration.}
\end{figure}

We now consider the total magnetic energy, $U (\mathbf{X}, \bm{\xi})$, which can be expressed to lowest order as $U = U_0 + U_\mathrm{p}(\mathbf{X}) + U_\mathrm{d}(\bm{\xi})$, where $U_0$ is the ground state energy, and $U_\mathrm{p}(\mathbf{X})$ and $U_\mathrm{d}(\bm{\xi})$ are the potential and deformation-induced energies, respectively. In a thin film disk, $U_\mathrm{p}(\mathbf{X}) \approx (\kappa_\mathrm{p}/ 2) \| \mathbf{X} \|^2$ to lowest order in $\mathbf{X}$, where $\kappa_\mathrm{p}$ is a stiffness coefficient~\cite{Guslienko:2001hz, Guslienko2002}. Since the deformation energy is an even function in $\bm{\xi}$, we can also write to lowest order in $\bm{\xi}$,
\begin{equation} \label{Eq:Ud}
	U_\mathrm{d}(\bm{\xi}) \approx \frac{1}{2}\kappa_\mathrm{d} \| \bm{\xi} \|^2.
\end{equation}
The parameter $\kappa_\mathrm{d}$ can be determined numerically from micromagnetics simulations. We consider a cylindrical dot with a diameter of $2R = 512$ nm and a thickness of $d = 20$ nm. We use magnetic parameters corresponding to permalloy, where the exchange constant is taken to be $A_\mathrm{ex} = 10$ pJ/m, $M_\mathrm{s} = 0.8$ MA/m, and $\alpha = 0.013$. We consider small deformations of the order $\|\bm{\xi}\| < 0.1 \text{ nm}$.  From the initial vortex state, which is found from energy minimization, we obtain the deformed configuration using Eq.~(\ref{Eq:config}), then calculate both the exchange and magnetostatic energies. $U_\mathrm{d}$ is plotted in Fig.~\ref{Fig:kd}(a) as a function of $\| \bm{\xi} \|^2$, where the slope corresponds to $\kappa_\mathrm{d} / 2$. For this set of parameters, $\kappa_\mathrm{d} \approx 0.14$ J/m$^{2}$.

Using the fitted value of $\kappa_\mathrm{d}$, we then calculate inertial motion of the core under a static in-plane magnetic field, $\|\mathbf{B}\| = 2$ mT. For this, we include an additional Zeeman term, $U_\mathrm{B} (\mathbf{X}) = \mu \left( \hat{\mathbf{z}} \times \mathbf{B} \right) \cdot \mathbf{X}$, where $\mu \approx (2/3) M_\mathrm{s} / R$ \cite{Guslienko2006}. The numerically calculated core trajectory and velocity using Eq.~(\ref{Eq:extThiele}) with $\kappa_\mathrm{d} = 0.14 \text{ J}/\text{m}^{2}$ are presented in Figs.~\ref{Fig:kd}(b) and \ref{Fig:kd}(c), respectively. After applying the field, the core exhibits additional oscillatory motion on top of the usual gyration, which also results in oscillations in the speed $\| \dot{\mathbf{X}} \|$, which are in good agreement with micromagnetic simulations as shown in Figs.~\ref{Fig:kd}(d) and \ref{Fig:kd}(e). These oscillations results from the core deformation. Note that we can observe an additional oscillation in the envelope of $\|\dot{\mathbf{X}}\|$ in the simulation [Fig.~\ref{Fig:kd}(e)], because the frequency of this oscillation $\sim 8 \text{ GHz}$ is close to the frequency of spin wave modes of this system.

By considering a linear steady-state motion of the core, the relation between the deformation energy and the core velocity, $U_\mathrm{d} = \frac{1}{2} \mathcal{M} \| \dot{\mathbf{X}} \|^2$, can be obtained from Eqs.~(\ref{Eq:extThiele}) and (\ref{Eq:Ud}), where
\begin{equation} \label{Eq:mass}
	\begin{split}
    	\mathcal{M} \equiv \frac{D^2}{\kappa_\mathrm{d}}
    \end{split}
\end{equation}
which can be regarded as an effective inertial mass under small deformations. In the case of $2R = 512$ nm and $d = 20$ nm of a NiFe cylindrical dot, for example, the mass is found to be $\mathcal{M}_\mathrm{NiFe} \approx 1.4 \times 10^{-23}$ kg, which is similar to the values for domain walls and bubbles obtained from experiments~\cite{Saitoh2004, Buttner2015}.

The acceleration of the core can be calculated from $\ddot{\mathbf{X}} = \mathbf{F}/\mathcal{M}$, where $\mathbf{F}$ is a force exerted on the core. If $\| \mathbf{B} \| = 2 \text{ mT}$ is applied on a stable vortex core, the acceleration will be $\| \ddot{\mathbf{X}} \| = \| \mathbf{F}_\mathrm{B} \|/\mathcal{M}_\mathrm{NiFe} \approx 1.2 \times 10^{12} \text{ m} / \text{s}^2$, where $\mathbf{F}_\mathrm{B} = -\mu |\mathbf{B}| \pi R^2 d$ is the force from the in-plane field \cite{Guslienko2006}. As shown in Figs.~\ref{Fig:kd}(c) and \ref{Fig:kd}(e) [dashed line], the calculated acceleration value is in a good agreement with those obtained from the simulation $ \|\ddot{\mathbf{X}} \| \approx 1.1 \times 10^{12} \text{ m} / \text{s}^2$.


%
\begin{figure}[t!]
\centering\includegraphics[width=6.7cm]{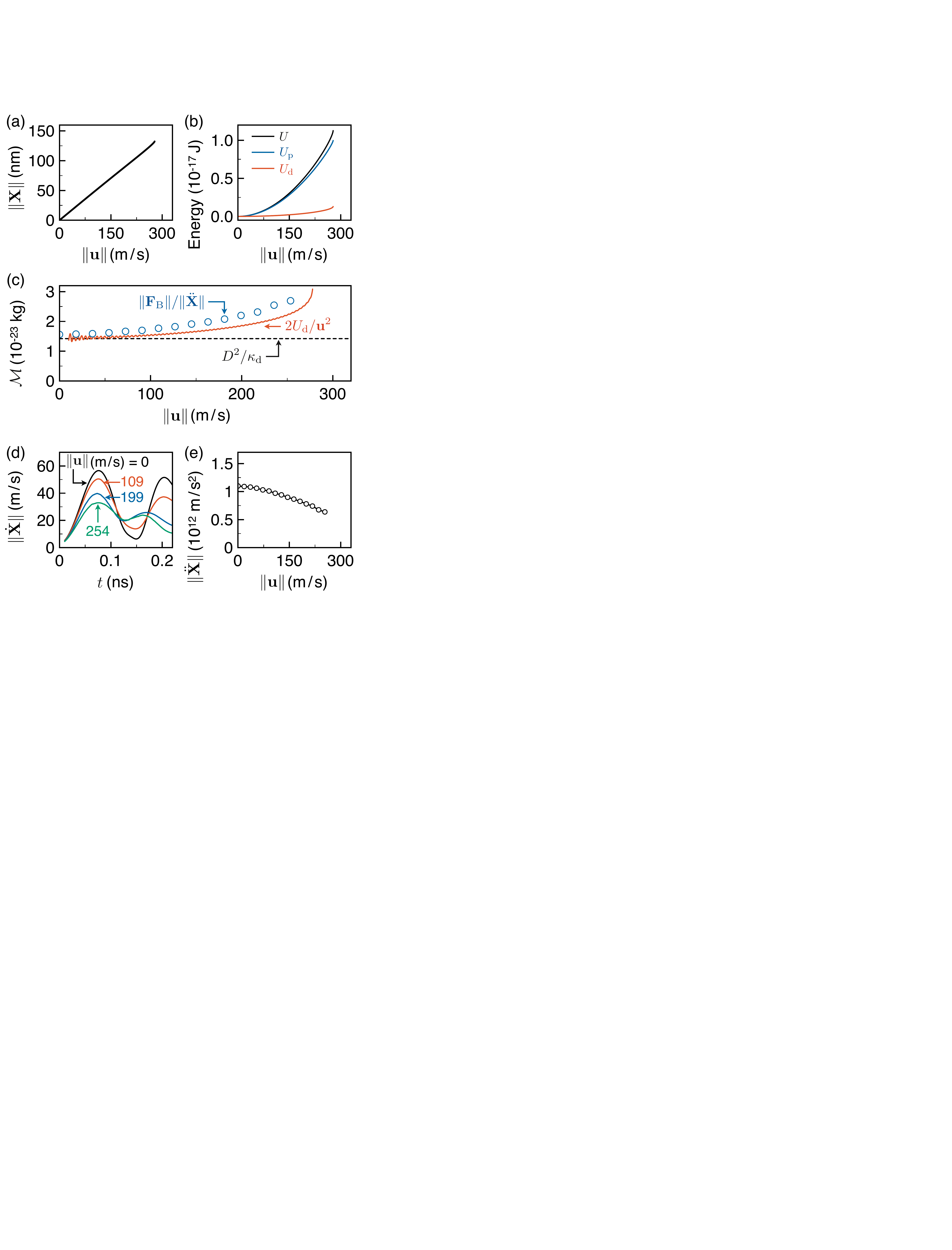} 
\caption{\label{Fig:mass}
    (a) Core position displacement, $\|\mathbf{X}\|$, as a function of an effective spin-current drift velocity, $\|\mathbf{u}\|$. (b) Total magnetic energy, $U$, the potential energy, $U_\mathrm{p}$, and the deformation energy, $U_\mathrm{d}$, as a function of $\|\mathbf{u}\|$. (c) Inertial mass of the vortex, $\mathcal{M}$, as a function of $\|\mathbf{u}\|$ obtained from $\mathcal{M} = 2 U_\mathrm{d}/\mathbf{u}^2$ (red line). The black dashed line indicates the mass at rest calculated from Eq.~(\ref{Eq:mass}). The blue circles are $\mathcal{M}$ obtained from the acceleration, $\|\ddot{\mathbf{X}}\|$, shown in (e). (d) Time evolution of $\|\dot{\mathbf{X}}\|$ for different $\|\mathbf{u}\|$. (e) The initial acceleration, $\|\ddot{\mathbf{X}}\|$, obtained from (d). 
   }
\end{figure}

We now turn our attention to the inertial mass under large deformations beyond the range for which Eqs.~(\ref{Eq:Ud}) and (\ref{Eq:mass}) are valid. In this limit, the deformation model  [Eq.~(\ref{Eq:config})] cannot be used to calculate $U_\mathrm{d}$ since the magnetic energy is overestimated particularly at the core center~\cite{SM}. Therefore, we extract the inertial mass from micromagnetic simulations at stable deformed-vortex configurations under a static-uniform in-plane electric current~\cite{Kravchuk2011}. Under applied currents, deformed core configurations can be obtained without any core motion, since the effective spin-current drift velocity $\mathbf{u}$ acts as the core-motion velocity $-\dot{\mathbf{X}}$ when only adiabatic term is considered in the limit $\alpha \ll 1$. In the immobile state, we obtain the core displacement from the center of the disk, $\| \mathbf{X} \|$, and the total magnetic energy, $U$, as a function of $\| \mathbf{u} \|$ [black lines in Figs.~\ref{Fig:mass}(a) and \ref{Fig:mass}(b)]; both $\| \mathbf{X} \|$ and $U$ increase with an increasing $\| \mathbf{u} \|$ up to the critical value at $\sim280 \text{ m} / \text{s}$ above which the stationary state changes due to reversal of the vortex core.

From the $\| \mathbf{X} \|$ and $U$ obtained, we dissociate the deformation energy, $U_\mathrm{d} = U - U_0 - (\kappa_\mathrm{p} / 2) \| \mathbf{X} \|^2$ [red line in Fig.~\ref{Fig:mass}(b)], then calculate the effective inertial mass from $\mathcal{M} = 2 U_\mathrm{d}/ \| \mathbf{u} \|^2$. Here we assume that $\mathcal{M} = D^2/\kappa_\mathrm{d}$ when $\| \mathbf{u} \| = 0$ to find the reasonable $\kappa_\mathrm{p}$ value. The obtained $\mathcal{M}$ is plotted in Fig.~\ref{Fig:mass}(c) as a function of $\| \mathbf{u} \|$ (red line). The result shows that the mass depends on the velocity; $\mathcal{M}$ gradually increases with an increasing $\|\mathbf{u}\|$ from $D^2/\kappa_\mathrm{d}$, in particular, $\mathcal{M}$ increases drastically near the critical velocity, $\sim 280 \text{ m} / \text{s}$. Because $\|\mathbf{u}\| \approx \|\dot{\mathbf{X}} \|$ in this case, the result implies that the effective inertial mass increases with an increasing the core velocity.

The core acceleration, $\|\ddot{\mathbf{X}}\|$, can be obtained using micromagnetic simulations in which an in-plane field, $\|\mathbf{B}\| = 2 \text{ mT}$, is applied to different stationary deformed configurations under $\|\mathbf{u}\|$. We plot the time evolution of the velocity in Fig.~\ref{Fig:mass}(d) and present their initial acceleration as a function of $\|\mathbf{u}\|$ in Fig.~\ref{Fig:mass}(e). The result shows that the acceleration reduces with increasing $\|\mathbf{u}\|$ as expected from the increase in mass. By using the obtained $\|\ddot{\mathbf{X}}\|$, we calculate the inertial mass from $\mathcal{M} = \|\mathbf{F}_\mathrm{B}\| / \|\ddot{\mathbf{X}}\|$ and plot it in Fig.~\ref{Fig:mass}(c) using blue circles, which is in good agreement with the mass calculated for the deformation energy. This result shows that the effective mass depends on the core velocity and the core deformation.

\begin{figure}
\centering\includegraphics[width=7.4cm]{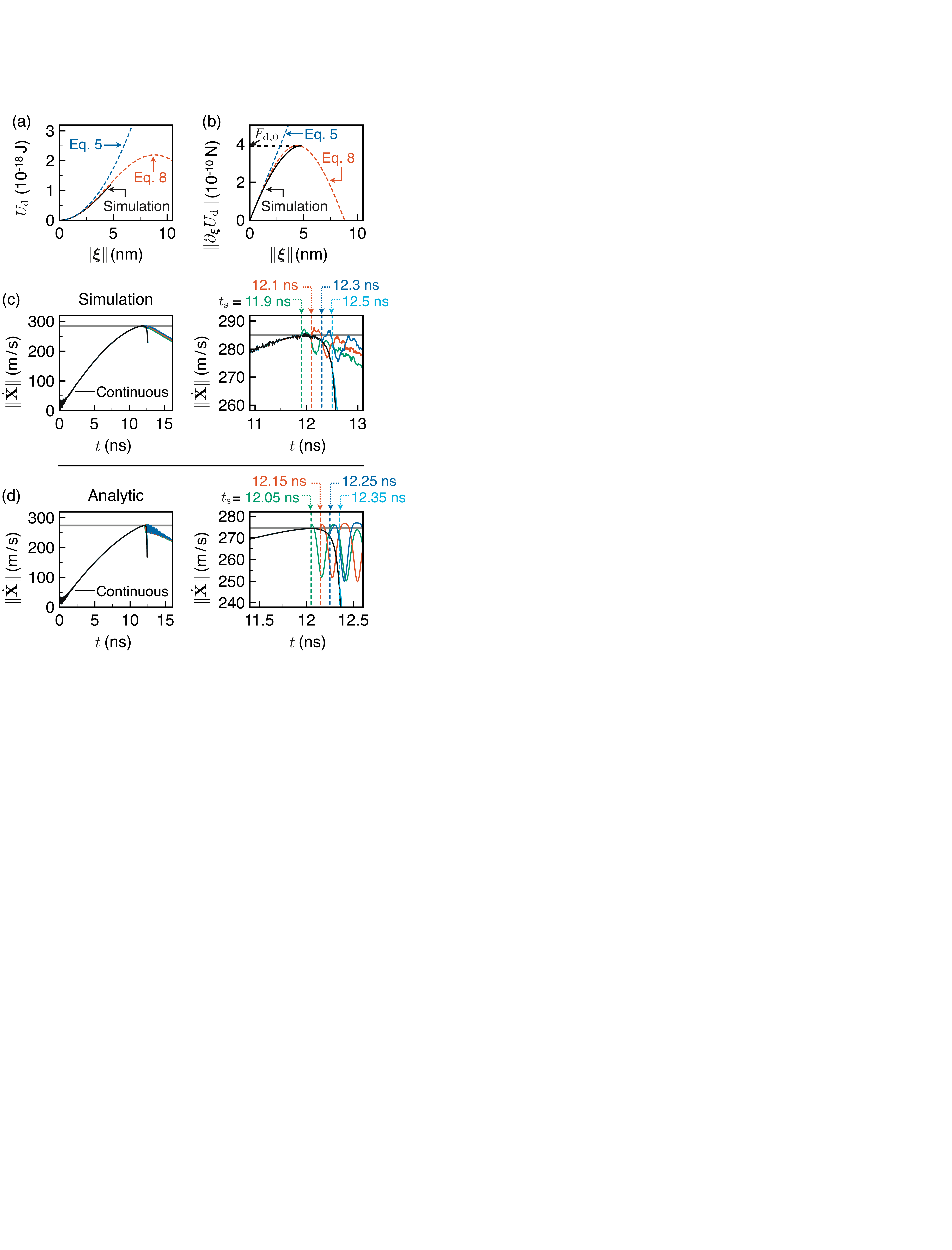} 
\caption{\label{Fig:Energy_Profile}
	(a) Deformation energy, $U_\mathrm{d}$, as a function of the deformation amplitude, $\|\bm{\xi}\|$ (black line). The blue dashed line is Eq.~(\ref{Eq:Ud}) and the red dashed line is Eq.~(\ref{Eq:ProfileModel}). (b) Derivative of the deformation energy, $\|\partial_{\bm{\xi}} U_\mathrm{d}\|$, as a function of $\|\bm{\xi}\|$. (c) $\|\dot{\mathbf{X}}\|$ during the resonant excitation obtained from micromagnetic simulations. The black line is the result from a continuous excitation up to the core switching and the gray line indicates the maximum core velocity. The green, red, blue, and light blue lines indicate the velocity when the driving field is turned off at $t_\mathrm{s}$ as shown in the right panel which shows a zoom on the region around the maximum core velocity. (d) as in (c) but obtained by solving  Eq.~(\ref{Eq:extThiele}) with Eq.~(\ref{Eq:ProfileModel}).}
\end{figure}

The increase of the effective inertial mass means that Eq.~(\ref{Eq:Ud}) is not valid for large deformations. Nevertheless, we can extract the relation between $U_\mathrm{d}$ and $\bm{\xi}$ from micromagnetic simulations. The equations of motion with an adiabatic spin-transfer torque can be calculated by using the substitution $d/dt \to d/dt + \mathbf{u} \cdot \nabla$ in the equations of motion,
\begin{subequations}
\begin{align} 
	\mathbf{G} \times \left( \dot{\mathbf{X}}-\mathbf{u} - \alpha \dot{\bm{\xi}} \right) + \tensor{D} \cdot \left(\alpha \dot{\mathbf{X}} + \dot{\bm{\xi}}\right) &= -\frac{\partial U}{\partial \mathbf{X}}; \\
	\mathbf{G} \times \left(  \dot{\bm{\xi}} + \alpha \dot{\mathbf{X}} \right)  + \tensor{D} \cdot \left(\alpha  \dot{\bm{\xi}}  -  \dot{\mathbf{X}} +  \mathbf{u}\right) &= -\frac{\partial U}{\partial \bm{\xi}}.
\end{align}
\label{Eq:extThieleAd}
\end{subequations}
From the Eq.~(\ref{Eq:extThieleAd}b), we obtain a relation $D u_x = -\partial U_\mathrm{d} / \partial \xi_x$ at the stationary deformed state under a $x$-directional in-plane current stated above, because $\dot{X} = \dot{Y} = \dot{\xi_x} = \dot{\xi_y} = 0$. $D$, $u_x$, and $U_\mathrm{d}$ values can be taken from the simulations, thus we can numerically obtain $\xi_x$ by assuming $\partial U_\mathrm{d} = \Delta U_\mathrm{d}$ and $\partial \xi_x = \Delta \xi_x$. The computed deformation energy, $U_\mathrm{d}$, and its derivative $\|\partial_{\bm{\xi}} U_\mathrm{d}\|$ are shown as a function of $\|\bm{\xi}\|$ in Figs.~\ref{Fig:Energy_Profile}(a) and \ref{Fig:Energy_Profile}(b) [black lines], respectively. For comparison, we also show Eq.~(\ref{Eq:Ud}) using blue dashed lines. The results illustrate that the obtained $U_\mathrm{d}$ obeys Eq.~(\ref{Eq:Ud}) for a small deformation, however, deviations appear when $\|\bm{\xi}\|$ becomes large. In particular, the slope of $\|\partial_{\bm{\xi}} U_\mathrm{d}\|$ almost vanishes at the maximum value, $F_\mathrm{d,0}$. This energy profile is consistent with the increase of $\mathcal{M}$, because the effective $\kappa_\mathrm{d}$ decreases when the deformation is large.

Based on the results in Figs.~\ref{Fig:Energy_Profile}(a) and \ref{Fig:Energy_Profile}(b), and by analogy with the Walker breakdown problem in domain wall dynamics, we propose a simple model for the deformation energy profile in place of Eq.~(\ref{Eq:Ud}),
\begin{equation} \label{Eq:ProfileModel}
	U_\mathrm{d} = U_\mathrm{d,0} \sin^2 \left( \sqrt{\frac{\kappa_\mathrm{d}}{2 U_\mathrm{d,0}}} \|\bm{\xi}\| \right),
\end{equation}
which is valid for deformations up to $\|\bm{\xi}_\mathrm{max}\| = (\pi/4)\sqrt{2 U_\mathrm{d,0}/\kappa_\mathrm{d}}$, i.e., the point at which the restoring force attains its maximum. $U_\mathrm{d,0}$ can be calculated from $U_\mathrm{d,0} = 2 F_\mathrm{d,0}^2/ \kappa_\mathrm{d}$, where $F_\mathrm{d,0}$ is the maximum value of $\|\partial_{\bm{\xi}} U\|$ which can be obtained from the simulation as shown in Fig.~\ref{Fig:Energy_Profile}(b). We plot Eq.~(\ref{Eq:ProfileModel}) in Figs.~\ref{Fig:Energy_Profile}(a) and \ref{Fig:Energy_Profile}(b) (red dashed line) with $F_\mathrm{d,0} = 3.9 \times 10^{-10} \text{ N}$; Eq.~(\ref{Eq:ProfileModel}) well describes the energy profiles before reaching $F_\mathrm{d,0}$. This model explains the maximum core velocity attained before core switching. If the core motion is carefully excited with sufficiently small $\|\dot{\bm{\xi}}\|$, $U_\mathrm{d}$ increases following Eq.~\ref{Eq:ProfileModel} and the core switching occurs when $U_\mathrm{d} = U_\mathrm{d,0}$. During the dynamics, the core velocity is approximately $\|\dot{\mathbf{X}}\| \approx \|\partial_{\bm{\xi}} U\|/D$ if $\alpha \ll 1$, therefore, generally the velocity does not exceed the maximum value, $F_\mathrm{d,0}/D$. Note that the maximum velocity here is analogous with the Walker velocity in 1D domain wall dynamics.

Eq.~(\ref{Eq:ProfileModel}) also shows that the maximum velocity is not a sufficient condition for the core switching, since further deformation is required to reach the maximum energy $U_\mathrm{d,0}$. This process is not instantaneous, so if the driving force is switched off prior to the switching event, the core reversal will not occur even though the velocity maximum has been attained. The transient dynamics associated with this is shown in Fig.~\ref{Fig:Energy_Profile}(c). With the same material and geometrical parameters above, the core motion is resonantly excited by a rotating field, $\mathbf{B}_\mathrm{CCW} = (B_0 \cos \omega t, B_0 \sin \omega t, 0)$, where $B_0 = 1.3 \text{ mT}$ and $\omega/2\pi = 320$~MHz. When the field is applied continuously, the core reaches the maximum velocity at $\sim 11.9$~ns, after which the velocity decreases and the core reverses at $\sim 12.6 \text{ ns}$ [black line in Fig.~\ref{Fig:Energy_Profile}(c)]. If however we turn off the field just after the maximum velocity is reached, core reversal does not take place [green, red, and blue lines in Fig.~\ref{Fig:Energy_Profile}(c)]. In those cases, the core velocity can even exceed the nominal maximum velocity instantaneously just after turning off the field, because of a relatively large $\dot{\bm{\xi}}$ from the deformation dynamics. Core switching occurs if the field is turned off too late [light blue line in Fig.~\ref{Fig:Energy_Profile}(c)]. These phenomena can be reproduced in our model by solving Eq.~(\ref{Eq:extThiele}) with Eq.~(\ref{Eq:ProfileModel}), as shown in Fig.~\ref{Fig:Energy_Profile}(d). This result shows that the energy-profile model, Eq.~(\ref{Eq:ProfileModel}), can describe the dynamics up to large deformations close to the switching transition.

%

\begin{acknowledgments}
This work was partially supported by the Agence Nationale de la Recherche (France) under Contract No. ANR-17-CE24-0008 (CHIPMuNCS).
\end{acknowledgments}


%

\end{document}


\title{Supplementary Material: ``Analytical model of the inertial dynamics of a magnetic vortex''}

\author{Myoung-Woo Yoo}
\affiliation{\CNN}
\affiliation{\LMOPS}
\author{Francesca Mineo}
\affiliation{\CNN}
\affiliation{\MPG}
\author{Joo-Von Kim}
\affiliation{\CNN}

\date{\today}

\begin{abstract}
This document provides additional information on the derivation of the equations of motion from the deformation model, a discussion on the accuracy of estimates of the deformation energy, and details concerning the micromagnetics simulations performed.
\end{abstract}

\maketitle


\subsection{Equations of Motion}
From the deformation model,
\begin{equation}
\begin{split} 
		\theta(\mathbf{x}, \bm{\xi}) &= \theta_0 -\sin \theta_0 \left[ \bm{\xi} \cdot \bm{\nabla} \phi_0 \right]; \\
		\phi(\mathbf{x}, \bm{\xi}) &= \phi_0 +\frac{1}{\sin \theta_0} \left[ \bm{\xi} \cdot \bm{\nabla} \theta_0 \right],
\end{split}
\label{Eq:Deformation_Model}
\end{equation}
we can calculate the Euler-Lagrange equation~\cite{Slonczewski1972, Thiaville2006, Kim2012},
\begin{equation} \label{Eq:Lagrange}
\begin{split}
   \frac{d}{dt} \frac{\partial L}{\partial \dot{q}} - \frac{\partial L_\mathrm{B}}{\partial q} + \frac{\partial W}{\partial \dot{q}} &= 0,
\end{split}
\end{equation}
where $q = \{X, Y, \xi_x, \xi_y \}$ represent the generalized coordinates and
\begin{equation} \label{Eq:Density}
\begin{split}
    L &= \frac{M_\mathrm{s}}{\gamma} \dot{\phi} \left( 1 - \cos \theta \right) - U, \\
    W &= \frac{M_\mathrm{s}}{2 \gamma} \left[ \left(\frac{d \theta}{dt} \right)^2 + \sin^2 \theta \left(\frac{d \phi}{dt} \right)^2 \right],
\end{split}
\end{equation}
are the Lagrangian and dissipative function densities, respectively. By solving the Euler-Lagrange equation and integrating over the volume, we obtain the equations of motion,
\begin{equation}
\begin{split}
    	\alpha D_{xx}^* \dot{X} + (-G_{xy}^* + \alpha D_{xy}^*) \dot{Y} + (D_{xx}^\dagger + \alpha G_{xx}^\dagger ) \dot{\xi_x} + (D_{xy}^\dagger + \alpha G_{xy}^\dagger) \dot{\xi_y} &= -\frac{\partial U}{\partial X}; \\
    	(-G_{yx}^* + \alpha D_{yx}^*) \dot{X} + \alpha D_{yy}^* \dot{Y} + ( D_{yx}^\dagger -\alpha G_{yx}^\dagger) \dot{\xi_x} + (D_{yy}^\dagger + \alpha G_{yy}^\dagger) \dot{\xi_y} &= -\frac{\partial U}{\partial Y}; \\
    	(-D_{xx}^\dagger + \alpha G_{xx}^\dagger ) \dot{X} + (-D_{xy}^\dagger - \alpha G_{xy}^\dagger) \dot{Y} + \alpha D_{xx}^\mathsection \dot{\xi_x} + (-G_{xy}^\mathsection + \alpha D_{xy}^\mathsection) \dot{\xi_y} &= -\frac{\partial U}{\partial \xi_x}; \\
    	(-D_{yx}^\dagger - \alpha G_{yx}^\dagger ) \dot{X} + (-D_{yy}^\dagger + \alpha G_{yy}^\dagger) \dot{Y} + (-G_{yx}^\mathsection + \alpha D_{yx}^\mathsection) \dot{\xi_x} + \alpha D_{yy}^\mathsection \dot{\xi_y} &= -\frac{\partial U}{\partial \xi_y},
\end{split}
\label{Eq:Thiele_Full}
\end{equation}
where
\begin{equation}
\begin{split} 
	G_{ij}^* &= \frac{M_\mathrm{s}}{\gamma} \int dV \sin \theta \left( \frac{\partial \theta}{\partial i}\frac{\partial \phi}{\partial j} - \frac{\partial \phi}{\partial i} \frac{\partial \theta}{\partial j} \right); \\
	G_{ij}^\dagger &= \frac{M_\mathrm{s}}{\gamma} \int dV \left( \sin \theta_0 \frac{\partial \theta}{\partial i}\frac{\partial \phi_0}{\partial j} - \frac{\sin^2 \theta}{\sin \theta_0} \frac{\partial \phi}{\partial i} \frac{\partial \theta_0}{\partial j} \right); \\
	G_{ij}^\mathsection &= \frac{M_\mathrm{s}}{\gamma} \int dV \sin \theta \left( \frac{\partial \theta_0}{\partial i}\frac{\partial \phi_0}{\partial j} - \frac{\partial \phi_0}{\partial i} \frac{\partial \theta_0}{\partial j} \right); \\
	D_{ij}^* &= \frac{M_\mathrm{s}}{\gamma} \int dV \left( \frac{\partial \theta}{\partial i} \frac{\partial \theta}{\partial j} + \sin^2 \theta \frac{\partial \phi}{\partial i} \frac{\partial \phi}{\partial j} \right); \\
	D_{ij}^\dagger &= \frac{M_\mathrm{s}}{\gamma} \int dV \left( \frac{\sin \theta}{\sin \theta_0} \frac{\partial \theta}{\partial i} \frac{\partial \theta_0}{\partial j} + \sin \theta \sin \theta_0 \frac{\partial \phi}{\partial i} \frac{\partial \phi_0}{\partial j} \right); \\
	D_{ij}^\mathsection &= \frac{M_\mathrm{s}}{\gamma} \int dV \left( \frac{\sin^2 \theta}{\sin^2 \theta_0} \frac{\partial \theta_0}{\partial i} \frac{\partial \theta_0}{\partial j} + \sin^2 \theta_0 \frac{\partial \phi_0}{\partial i} \frac{\partial \phi_0}{\partial j} \right).
\end{split}
\label{Eq:Thiele_Parameters}
\end{equation}
In the case of a static vortex configuration ($\|\bm{\xi}\| = 0$), we obtain
\begin{equation}
\begin{split} 
	G_{xx}^* &= G_{xx}^\dagger  = G_{xx}^\mathsection = G_{yy}^* = G_{yy}^\dagger  = G_{yy}^\mathsection = 0; \\
	G_{xy}^* &= G_{xy}^\dagger = G_{xy}^\mathsection = -G_{yx}^* = -G_{yx}^\dagger = -G_{yx}^\mathsection = G; \\
	D_{xx}^* &= D_{xx}^\dagger = D_{xx}^\mathsection = D_{yy}^* = D_{yy}^\dagger = D_{yy}^\mathsection = D; \\
	D_{xy}^* &= D_{xy}^\dagger = D_{xy}^\mathsection = D_{yx}^* = D_{yx}^\dagger = D_{yx}^\mathsection = 0,
\end{split}
\label{Eq:Thiele_Parameters_0}
\end{equation}
where
\begin{equation}
\begin{split} 
	G &= \frac{M_\mathrm{s}}{\gamma} \int dV \sin \theta_0 \left( \frac{\partial \theta_0}{\partial x}\frac{\partial \phi_0}{\partial y} - \frac{\partial \phi_0}{\partial y} \frac{\partial \theta_0}{\partial x} \right) = \frac{2 \pi M_\mathrm{s} d p } {\gamma}; \\
	D &= \frac{M_\mathrm{s}}{\gamma} \int dV \left( \frac{\partial \theta_0}{\partial x} \frac{\partial \theta_0}{\partial x} + \sin^2 \theta \frac{\partial \phi_0}{\partial x} \frac{\partial \phi_0}{\partial x} \right), \\
	&= \frac{M_\mathrm{s}}{\gamma} \int dV \left( \frac{\partial \theta_0}{\partial y} \frac{\partial \theta_0}{\partial y} + \sin^2 \theta \frac{\partial \phi_0}{\partial y} \frac{\partial \phi_0}{\partial y} \right), \\
	&\approx \frac{\pi M_\mathrm{s} d}{\gamma} \left[ 2 + \ln \left(\frac{R}{b} \right) \right].
\end{split}
\label{Eq:G0_and_D0}
\end{equation}
based on the Usov model~\cite{Guslienko2006}. By substituting Eq.~(\ref{Eq:Thiele_Parameters_0}) into Eq.~(\ref{Eq:Thiele_Full}), we obtain the equations of motion of the vortex core presented in the main text [Eq.~(4)].
%
\subsection{Overestimation of deformation energy}
\begin{figure}[htb]
\centering\includegraphics[width = 0.7\textwidth]{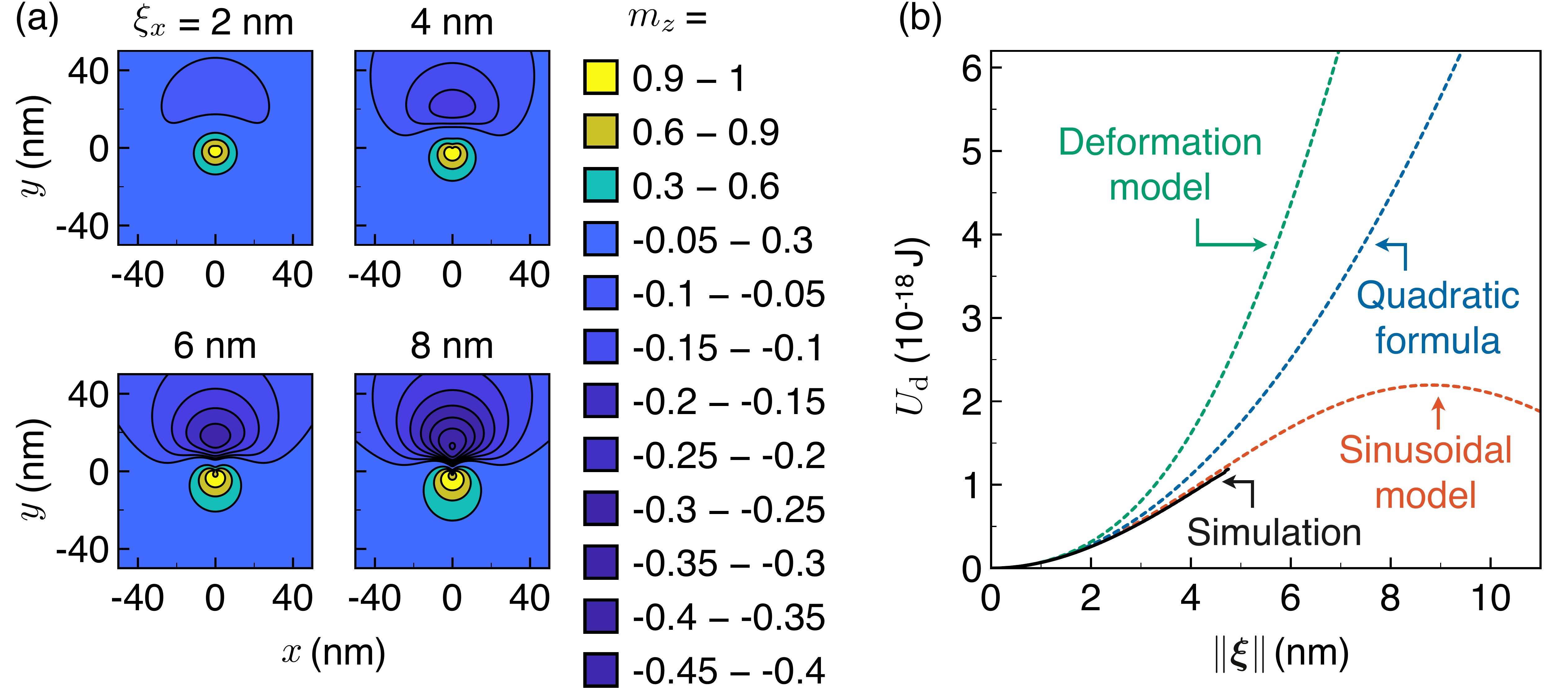} 
\caption{\label{Fig:Deformation}
  (a) Contour lines of $z$-component of the local magnetization, $m_z$, of deformed vortex-core configurations calculated from the analytic deformation model [Eq. ~(\ref{Eq:Deformation_Model})] with different deformation parameters, $\xi_x$. (b) $U_\mathrm{d}$ as a function of $\|\bm{\xi}\|$ obtained from the energy of the deformation model [Eq.~(\ref{Eq:Deformation_Model})] (green dashed line), quadratic model [Eq. (5) in the main text] (blue dashed line), sinusoidal model [Eq. (8) in the main text] (red dashed line), and a simulation [black line].
   }
\end{figure}

For a small deformation, the stiffness of the deformation energy, $\kappa_\mathrm{d}$, can be obtained by calculating energy from the deformation model [Eq.~(\ref{Eq:Deformation_Model})], as shown in Fig.~2(a) in the main text. However, the deformation energy, $U_\mathrm{d}$ can be overestimated for a large deformation, because the model cannot describe deformation of the vortex core properly [Fig.~\ref{Fig:Deformation}(a)]. This peculiar magnetic configuration results in an energy overestimation as shown in Fig.~\ref{Fig:Deformation}(b).

\subsection{Micromagnetic Simulations}

The MuMax3 code is used for micromagnetic simulations~\cite{Vansteenkiste2014}. A cylindrical disk is chosen with a diameter of $2R = 512 \text{ nm}$ and a thickness of $d = 20 \text{ nm}$. We use magnetic parameters of permalloy; we consider an exchange constant of $A_\mathrm{ex} = 10$ pJ/m, a saturation magnetization $M_\mathrm{s} = 0.8$ MA/m, and a Gilbert constant of $\alpha = 0.013$. For the potential and deformation energies, we used $\kappa_\mathrm{p} = 1.15 \times 10^{-3} \text{ J}/\text{m}^2$ and $\kappa_\mathrm{d} = 1.40 \times 10^{-1} \text{ J}/\text{m}^2$, respectively. The initial vortex state has clockwise chirality and upward core. To obtain dynamics of the vortex core, the system is uniformly discretized by a unit cell of $0.5 \times 0.5 \times 20 \text{ nm}^3$. To obtain stationary states with an in-plane current, $2.0 \times 2.0 \times 20 \text{ nm}^3$ unit cell is used. The in-plane current is applied in the $x$-direction with a spin polarization of $P = 0.5$. The non-adiabatic effect is ignored.

%

%